\documentclass[letter]{aa}  
\usepackage{graphicx}

\usepackage{txfonts}
\usepackage{xcolor}
\usepackage{upgreek}
\usepackage{hyperref}
\usepackage[normalem]{ulem}
\definecolor{ao(english)}{rgb}{0.0, 0.5, 0.0}

\usepackage{amstext}

\begin{document} 

   \title{Detection of the magnetar XTE$\,$J1810$-$197 at 150 and 260$\,$GHz\\with the NIKA2 kinetic inductance detector camera}

   \author{
   P.~Torne\inst{1,2}\thanks{Main contact author, \email{torne@iram.es} } \and
   J.~Mac\'ias-P\'erez\inst{3} \and
   B.~Ladjelate\inst{1}\and
   A.~Ritacco\inst{1,4,5} \and
   M.~S\'anchez-Portal\inst{1} \and
   S.~Berta\inst{6} \and
   G.~Paubert\inst{1} \and
   M.~Calvo\inst{7}\and
   G.~Desvignes\inst{8,9} \and
   R.~Karuppusamy\inst{9} \and
   S.~Navarro\inst{1} \and
   D.~John\inst{1}\and
   S.~S\'anchez\inst{1}\and
   J.~Pe\~nalver\inst{1}\and
   M.~Kramer\inst{9,10}\and
   K.~Schuster\inst{6}
   }

  \institute{
  Instituto de Radioastronom\'ia Milim\'etrica (IRAM), Avda. Divina Pastora 7, Local 20, 18012 Granada, Spain
  \and
  East Asian Observatory, 660 N. A’ohoku Place, University Park, Hilo, Hawaii 96720, USA
  \and
  Laboratoire de Physique Subatomique et de Cosmologie, Université Grenoble-Alpes, CNRS, 53 av. des Martyrs, 38026 Grenoble, France
  \and
  Institut d'Astrophysique Spatiale (IAS), CNRS and Université Paris-Sud, B\^atiment 121, Orsay, France
  \and 
  Laboratoire de Physique de l'\'Ecole Normale Sup\'erieure, ENS, 24 rue Lhomond, 75005 Paris, France
  \and
  Institut de Radioastronomie Millimétrique (IRAM), 300 rue de la Piscine, 38406 St. Martin d’Hères, France
  \and
  Institut N\'eel, CNRS and Universit\'e Grenoble-Alpes, 25 rue des Martyrs BP 166, 38042 Grenoble, France
  \and 
  LESIA, Observatoire de Paris, Université PSL, CNRS, Sorbonne Université, Université de Paris, 5 place Jules Janssen, 92195 Meudon, France
  \and
  Max-Planck-Institut f\"{u}r Radioastronomie, Auf dem H\"{u}gel 69, D-53121, Bonn, Germany
  \and
  Jodrell Bank Centre for Astrophysics, School of Physics and Astronomy, University of Manchester, Manchester M13 9PL, UK
  }

  \date{Received May 27, 2020; accepted July 3 , 2020}

 
  \abstract
   {The investigation of pulsars between millimetre and optical wavelengths is challenging due to the faintness of the pulsar signals and the relative low sensitivity of the available facilities compared to 100-m class telescopes operating in the centimetre band. The kinetic inductance detector (KID) technology offers large instantaneous bandwidths and a high sensitivity that can help to increase the ability of existing observatories at short wavelengths substantially to detect pulsars and transient emission.}
   {To investigate whether pulsars can be detected with KIDs, we observed the anomalous X-ray pulsar XTE~J1810$-$197 with the New IRAM KIDs Array-2 (NIKA2) camera installed at the IRAM 30m telescope in Spain.}
   {Several short observations of XTE$\,$J1810$-$197 were made on 2019 March 25 under good weather conditions to verify the stability of the KIDs and to try to detect the expected broadband pulsations from the neutron star.}
   {We detected the pulsations from the pulsar with NIKA2 at its two operating frequency bands, 150 and 260$\,$GHz ($\uplambda$=2.0 and 1.15$\,$mm, respectively). This is the first time that a pulsar is detected with a receiver based on KID technology in the millimetre band. In addition, this is the first report of short millimetre emission from XTE$\,$J1810$-$197 after its reactivation in December 2018, and it is the first time that the source is detected at 260$\,$GHz, which gives us new insights into the radio emission process of the star.}
   {We demonstrate that KIDs can fulfil the technical requirements for detecting pulsed emission from neutron stars in the millimetre band. We show that the magnetar XTE$\,$J1810$-$197 is again emitting strong pulsations in the short millimetre band.}

   \keywords{Stars: magnetars -- (Stars:) pulsars: individual XTE~J1810$-$197 -- Instrumentation: detectors -- Astronomical instrumentation, methods and techniques -- Radiation mechanisms: non-thermal}

   \titlerunning{Magnetar XTE~J1810$-$197 at 150 and 260 GHz with NIKA2}
   \authorrunning{P.~Torne et al.}

   \maketitle


\section{Introduction}

Pulsars are rotating neutron stars that emit broadband-beamed emission that is detectable as pulses from Earth as their beams cross our line of sight. They act like cosmic clocks, and studying them enables high-precision astronomy \citep[e.g.][]{kra06, eat13, thank20}. Nevertheless, the mechanism of their coherent radio emission is still unknown \citep[for a review, see e.g.][]{melrose17}.

One way to better understand the coherent and incoherent components of pulsar emission and aid constraining theoretical models is to measure the spectrum of pulsars in the poorly explored window between the radio and optical regimes \citep[e.g.][]{mich78}. However, pulsars in general are steep spectral sources in the radio band, which means that their emission is extremely weak above a few gigaherz \citep{mar2000}. Additionally, with higher frequencies, the collecting areas and efficiency of the available telescopes tend to decrease, and instrumentation with a high enough time resolution to resolve the pulsations is rarely available. As a result, pulsar studies in the millimetre, infrared, and optical regimes are sparse and highly challenging.

To improve our ability to detect and study pulsars at short wavelengths, we need higher sensitivity. One solution is to use large-bandwidth and low-noise receivers. In recent years, a new technology based on kinetic inductance detectors (KIDs) has been used to fabricate sensitive large-array detectors \citep[for a review, see][]{maus18}.  KID arrays of thousands of pixels can be built and operated with the most advanced sensitivity \citep{2011Monfardini,mazin13,2014Catalano,2016Calvo,2020Perotto}.

Furthermore, KIDs are fast detectors, with typical time constants of about 100$\,\mu$s \citep{2016Catalano,2016Monfardini}, and they can be used at very high sampling rates up to few megaherz \citep{2010ApPhL..96z3511S, 2016Bourrion, 2019Fasano}. This makes KIDs a promising tool for investigating pulsars in unexplored sections of their spectrum and test models of pulsar magnetospheres.

The KID-based NIKA2 camera installed at the IRAM~30m telescope \citep{adam18} 
is a dual wide-band camera operating simultaneously at 150 and 260$\,$GHz ($\uplambda$=2.0 and 1.15$\,$mm) with 45$\,$GHz of bandwidth and a total of 2896 KIDs cooled down to about 150$\,$mK \citep{2016Calvo}. NIKA2 offers an excellent performance with typical sensitivities per detector of 9 and 30~mJy$\,$s$^{1/2}$ at 150 and 260$\,$GHz, respectively \citep{2020Perotto}. In the future, NIKA2 will also offer polarisation measurements in the 260 GHz band \citep{2020ritacco}. Currently, the maximum theoretical sampling frequency of NIKA2 is 500 Hz\footnote{This fast-sampling mode is pending commissioning.}.


The observations presented here focus on the anomalous \mbox{X-ray} pulsar XTE~J1810$-$197. This pulsar was identified as a magnetar in 2003 after a sudden increase in its X-ray luminosity \citep{ibra04}. Magnetars are a small family of pulsars whose high-energy luminosity cannot be accounted for solely by the spin-down energy, and they are thought to be powered in addition by magnetic field decay \citep[for a review on magnetar properties, see e.g.][]{kasbol17}. XTE$\,$J1810$-$197 was furthermore the first magnetar to exhibit radio emission \citep{halpern2005, cam06}, an uncommon property that only five of the known 23 known magnetars have shown to date. Interestingly, and in contrast to the majority of the pulsar population, radio magnetars share a tendency to maintain a flat or even inverted spectrum in the radio band \citep{lev10, cam07c, cam08, tor15}. Radio magnetars can therefore be very bright in the millimetre band \citep{tor17}.

XTE$\,$J1810$-$197 is highly variable, and its radio emission ceased about three years after the discovery of the radio pulsations \citep{cam16}. However, radio emission from the neutron star was re-detected in December 2018 \citep{lev19}, and follow-up campaigns at several wavelengths including X-ray \citep{gott2019} confirmed that the source entered a new period of intense radiative activity. Observations were also carried out with the Superconductor-Insulator-Superconductor (SIS) Eight Mixer Receiver \citep[EMIR,][]{car12} at the IRAM 30m telescope, confirming that the source was also bright in the millimetre band. These observations will be presented elsewhere, but they confirmed that XTE$\,$J1810$-$197 was a good candidate for trying to  observe it with NIKA2 and test whether the KIDs are able to detect broadband pulsations in the millimetre band.


\section{Dataset and methods}

\subsection{Observations}

 XTE$\,$J1810$-$197 was observed on 2019 March 25 with the NIKA2 camera at the IRAM~30m telescope under good weather conditions ($\rm {\tau_z}^{225\,GHz} = 0.29 $). The KID arrays were configured to sample at a standard frequency $\rm F_s = 23.84\,Hz$, which is enough to obtain 132 data points across one rotational period of the pulsar ($P\approx5.54\,$s). The observing mode was set to stare mode, in which the telescope continuously tracked the source at a fixed position on the three arrays of the camera. 

The resonance frequency of each detector in NIKA2 is proportional to the input signal, and signal variations translate into resonance shifts. During the observations, we measure the signal via the amplitude and phase shift (transfer function) of a specific microwave test tone for each detector \citep{2016Calvo}. Prior to an observation, the detectors and readout are tuned to adapt the response to the input background signal from the sky. 
To ensure a coherent time series, the data were acquired continuously during each full scan and were recorded into a single data file.

In total, we recorded four scans of XTE$\,$J1810$-$197, the first two with 10$\,$min duration each, and the last two with 15$\,$min duration each. The maximum integration time per scan was set to ensure efficient and manageable data processing. 

\subsection{Data processing}\label{sec:proc}

The exact position of the magnetar on the KID arrays was a priori only known to a certain precision. To search for the exact pixels on which the source was centred, we wrote as 32 bit float time series the values of reconstructed signal (in resonance frequency units) from a number of pixels around the estimated position, corresponding to an area on sky of about 1 arcmin$^2$. Because the pixel size and separation are different for the 2 and 1.15$\,$mm arrays, time series from 45 pixels (slightly larger area) were extracted for the 2$\,$mm array (called A2), and 77 and 84 pixels for the two arrays at 1.15$\,$mm (called A1 and A3, respectively). The difference in the number of pixels for the two 1.15$\,$mm arrays is due to a slightly larger number of non-valid pixels in array A3 \citep{2020Perotto}.

Next, we filtered the time series to subtract the atmospheric variations. We applied a running-fit filter consisting of a moving window of 4.5$\,$s. As the window moves, a first-order polynomial is fit and subtracted from the data. Each filtered time series (i.e. from each extracted KID pixel) was then directly folded by the routine \texttt{prepfold} from the pulsar analysis suite PRESTO\footnote{\url{https://www.cv.nrao.edu/~sransom/presto/}}. An ephemeris from \citet{cam16}, obtained from the pulsar database PSRCAT\footnote{\url{https://www.atnf.csiro.au/research/pulsar/psrcat/}} \citep{man05}, was used to predict the spin period at the observation epochs. The resulting pulsar candidate plots were manually inspected to identify the time series, and so the KID pixel of each array that contained the best signal in terms of signal-to-noise ratio at the spin period of XTE$\,$J1810$-$197. The KID pixels containing the highest-significance pulsations were KH097, KC016, and KP018 for the A1, A2, and A3 array, respectively. 

A running-fit (or running-median or mean) filtering tends to subtract part of the pulsar signal, which affects the intensity measurements. For this reason, we re-processed the data after the pixels that contained the magnetar signal were identified. In this re-processing, we used a so-called multi-pixel filter. This method exploits the advantage of having multiple beams on the sky and removes the atmospheric variations, but preserves the pulsations unchanged. For 150$\,$GHz, we selected pixels pointed on the sky between 2 and 3 half-power beam widths (HPBW, $\theta_{150}
\!\!\approx$17.7$\,$arcsec) around the magnetar. Because the detectors are less sensitive and the atmospheric fluctuations stronger, we selected pixels at 260\ GHz that pointed somewhat closer, between 1.25 and 2.5 HPBWs ($\theta_{260}\!\!\approx$11.2$\,$arcsec). From this subset, the pixels showing residual pulsations at any signal-to-noise ratio (as found with the running-fit filter) were flagged and excluded from the analysis. For the remaining pixels (17 pixels for A2, 14 pixels for A1 and A3), a running median with a window of 1.5$\,$s was applied, and the resulting running median vectors were averaged and subtracted from the time series centred on the magnetar. This was done separately for arrays A1, A2, and A3. The time series at 260$\,$GHz still showed some residual slow fluctuations after this procedure. In these cases, a final smoothing by a running median of window 5$\,$s was applied. This filtering method resulted in a clean average profile with a flat off-pulse region, and without the typical dips resulting from direct running median or running fit subtractions.

Finally, the time series cleaned with the multi-pixel filter were reformatted as SIGPROC\footnote{\url{http://sigproc.sourceforge.net}} filterbank-type files by adding a header containing the metadata of the observation. The binary formatting of the header was made with tools from SIGPYPROC\footnote{\url{https://github.com/ewanbarr/sigpyproc}}. To increase the total signal-to-noise ratio, the four individual scans were concatenated by letting \texttt{prepfold} pad the gaps in between scans. For the 260$\,$GHz frequency, arrays A1 and A3 receive half of the total power each. Before concatenation and folding, the time series from the orthogonal linearly polarised pixels KH097 and KP018 were therefore combined to produce the total intensity output.

\subsection{Calibration}

To convert into flux density units, and prior to the analysis presented in Sec.~\ref{sec:proc}, we scaled the time series by calibration factors obtained from observations of Uranus taken in September 2018 and March 2019 \citep[see][for details of the NIKA2 calibration]{2020Perotto}. These observations were also used to compute the relative sky position between detectors. A correction was applied to the intensity at 260$\,$GHz to account for a slight misalignment of the beams from pixels KH097 and KP018 with respect to the position on sky of pixel KC016. The distance between the beams at 150$\,$ and 260$\,$GHz for the pixels we used was 4.5$\,$arcsec. The correction includes a pointing offset of 2$\,$arcsec from the magnetar position. The beam misalignment plus pointing offset result in an increase by a factor of two for the 260 GHz intensity. The uncertainties are derived from the noise statistics, combined with a 6 and 10\%\ 1$\sigma$ error for the 150 and 260 GHz bands, respectively, that originates from the scaling to Jansky factors \citep{2020Perotto}, combined with a 10\%\ 1$\sigma$ error for the correction due to the misalignment of the beams.

   \begin{table}
      \caption[]{XTE~J1810$-$197 measured properties at the epoch of observation (2019 March 25): MJD epoch, barycentered spin period, continuum-equivalent flux density at 150 and 260~GHz, and spectral index. Uncertainties in parentheses indicate the 1$\sigma$ error on the last significant quoted digit.}
         \label{tab:prop}
     $$ 
         \begin{array}{ccccc}
            \hline
            \noalign{\smallskip}
            \rm{Epoch}      &   P       & S_{150}   & S_{260} & \alpha \\
            \rm{(MJD)}      & (s) &   \rm{(mJy)}   &   \rm{(mJy)}\\
            \noalign{\smallskip}
            \hline
            \noalign{\smallskip}
            58567.23714 & 5.541396(3) & \rm 11.7(7)     & \rm 6.4(10)    & \rm -1.1(3) \\
            \noalign{\smallskip}
            \hline
         \end{array}
     $$ 
   \end{table}
%

%
   \begin{figure*}
   \centering
          \includegraphics[scale=0.73]{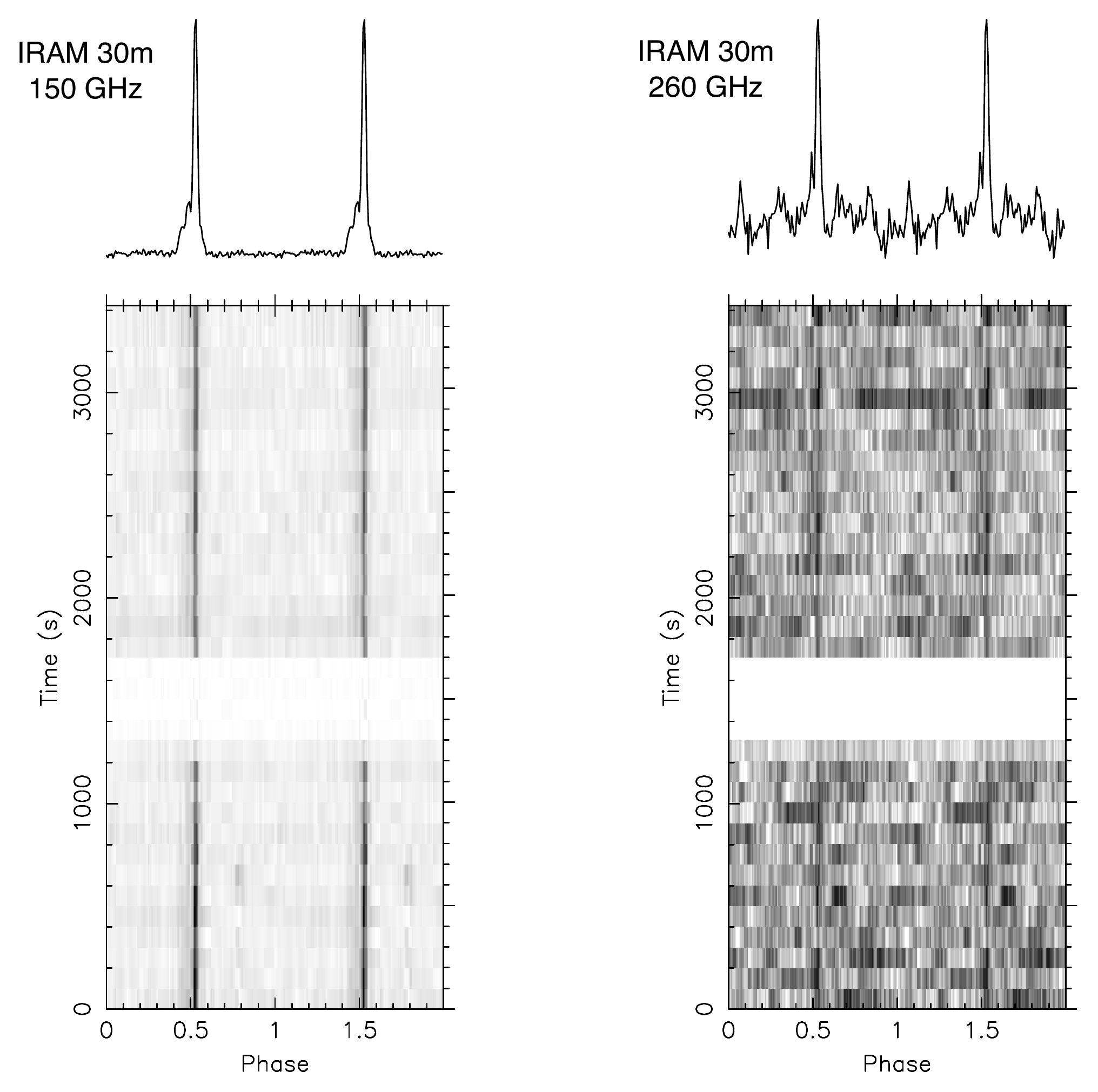}
          \caption{Average profiles (shown twice for clarity) of the detection of XTE~J1810$-$197 with NIKA2 at 150$\,$GHz (left) and 260~GHz (right). Profiles have 128 bins, corresponding to a time resolution of $\approx$43$\,$ms on the horizontal axis. The bottom panels show the signal intensity over time with 1.7 min resolution on the vertical axis. The white gap corresponds to the time when the telescope did not observe the source.}
          \label{fig:prestoprof2mm}
    \end{figure*}

%
   \begin{figure*}
   \centering
          \includegraphics[scale=0.65]{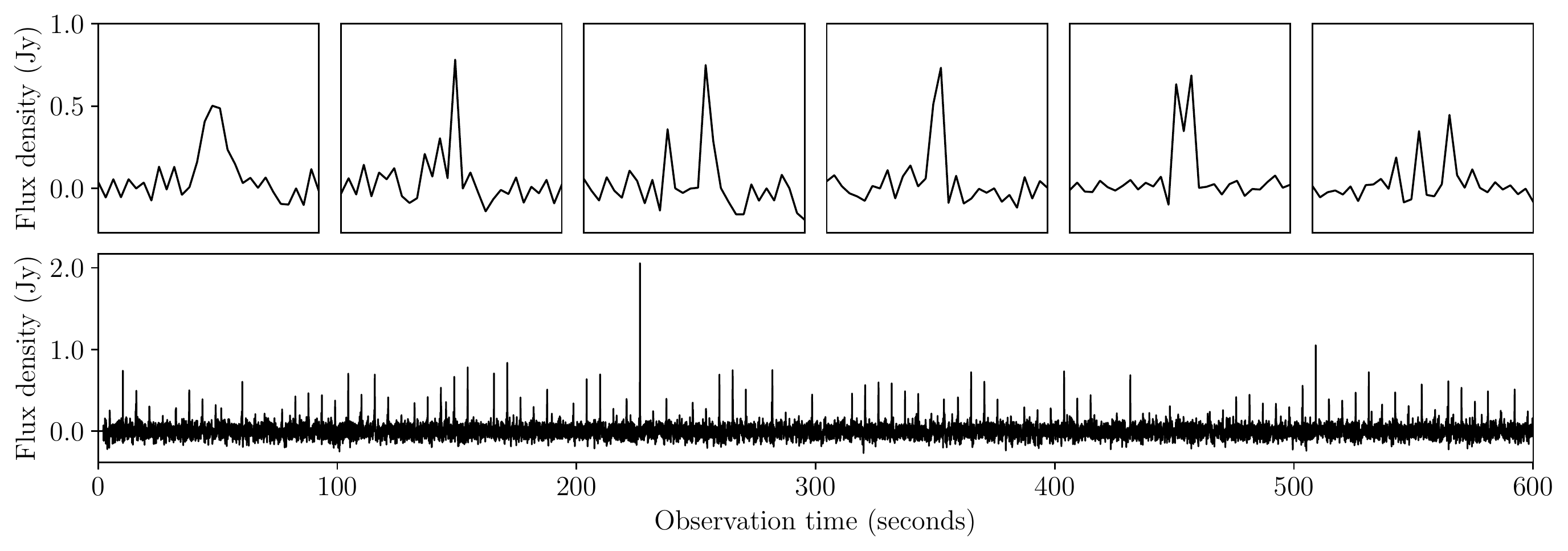}
          \caption{Single-pulse activity from XTE~J1810$-$197 during our observations with NIKA2 on 2019 March 25. Only the last 10$\,$min of the observation at 150~GHz are shown for clarity, but the activity is similar for the total 50$\,$min on-source. No single pulses are detected at 260~GHz, which may be an effect of spectral index and instrumental sensitivity. The upper panels show a zoom (1.25$\,$s around the peak) into a selection of six pulses, and they share the y-axis scale in Jansky units. The bottom panel shows the 10 min time series. The time resolution in all panels is $\approx$42$\,$ms, limited by the sampling frequency set for the KID arrays. Many pulses show structure. Most pulses are unresolved.}
          \label{fig:SPs}
    \end{figure*}

\section{Results and discussion}\label{sec:results}

The magnetar XTE$\,$J1810$-$197 was clearly detected in both bands of the NIKA2 camera at central frequencies 150 and 260$\,$GHz ($\uplambda$=2.0 and 1.15$\,$mm), with peak significances of 124$\sigma$ and 13$\sigma$, respectively. Figure \ref{fig:prestoprof2mm} shows the average profiles and signal intensities as a function of integration time. At 150~GHz we detect individual pulsations in almost every rotation of the magnetar (see Fig.~\ref{fig:SPs}).

%
   \begin{figure}
   \centering
          \includegraphics[scale=0.6]{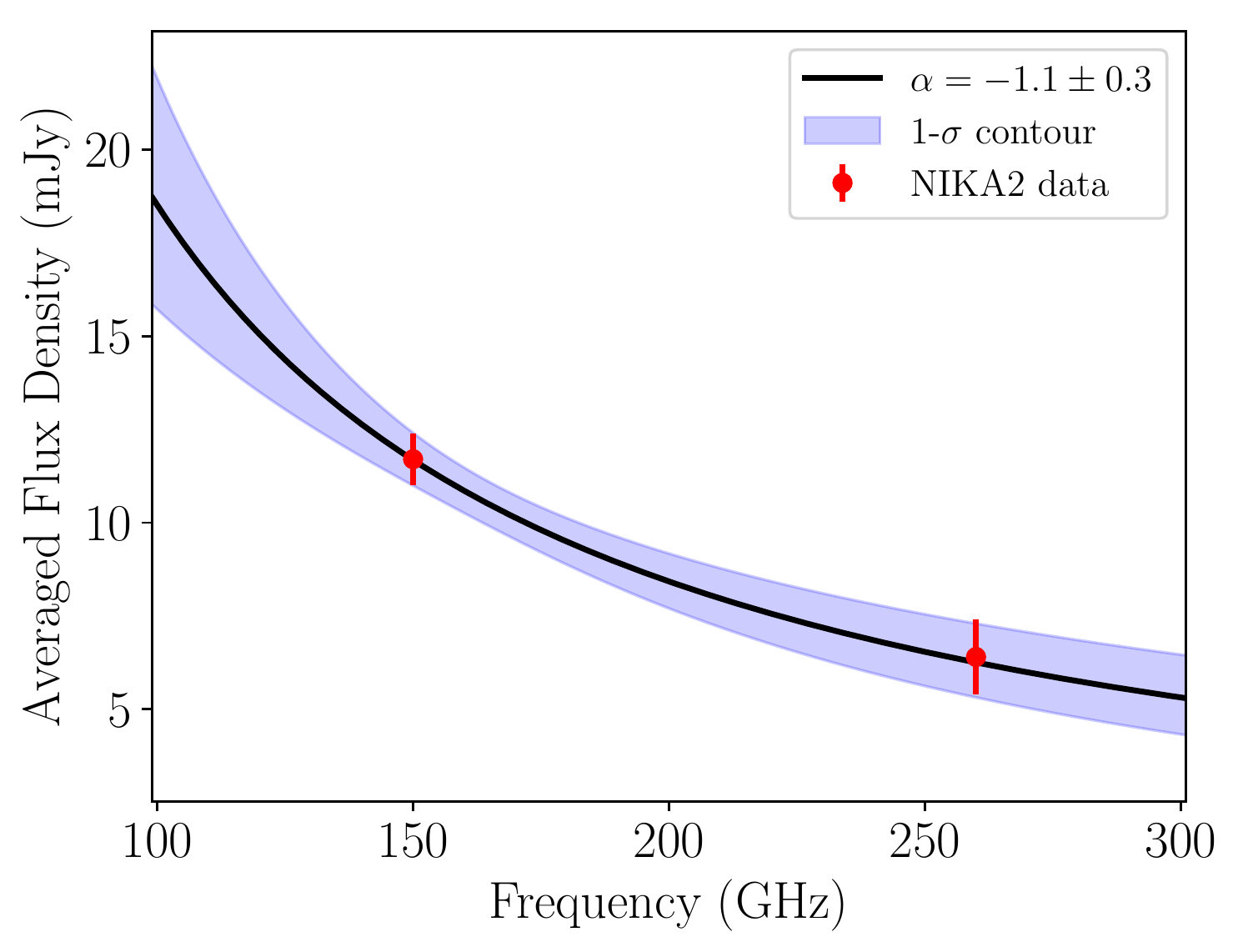}
          \caption{Emission spectrum in the millimetre band of the radio magnetar XTE~J1810$-$197 during the observations with NIKA2 on 2019 March 25. The solid black line indicates the mean value, and the shaded blue region encompasses the 68\% confidence interval. The spectral index is $\rm{\alpha}=-1.1\pm0.3$ (for $S_{\nu} \propto \nu^{\alpha}$).}
          \label{fig:spectrum}
    \end{figure}

Table~\ref{tab:prop} presents the continuum-equivalent (or averaged) flux density (i.e. the area below the pulse divided by the rotational period) and derived spectral index. The spectrum and fit are shown in Figure~\ref{fig:spectrum}. 
With $S_{150}=11.7\pm0.7\,$mJy and $S_{260}=6.4\pm1.0\,$mJy, the average spectral index is $\rm{\alpha}=-1.1\pm0.3$ (for $S_{\nu} \propto \nu^{\alpha}$). 
Because we have only a few data points and to better estimate the uncertainties, the spectral index was fit in a Markov chain Monte Carlo (MCMC) analysis using the Python package \texttt{emcee}\footnote{\url{https://emcee.readthedocs.io/}} \citep{2013ascl.soft03002F}.

The resulting spectrum is relatively flat, although much steeper than the inverted spectrum measured between 0.7$-$4.0$\,$GHz \citep[$\rm{\alpha}\approx+0.3$,][]{dai19}. The millimetre flux density is about $\text{nine}$ times higher than what can be estimated from the light curve reported by \citet{maan19} ($\approx$1.0$\,$mJy at 650$\,$MHz). When we include this low-frequency value, the spectral fit would result in $\rm{\alpha}\approx+0.39\pm0.02$, consistent with the derived value at lower frequency. The difference in spectral index resulting from using the millimetre data alone might be related to a poor constraint of the fit with only two data points. However, interstellar scintillation \citep[see the discussion in][]{laz08} and/or a frequency-dependent intrinsic variability of the magnetar emission may also play a role. We measured variations in the flux density of up to 50\% peak to peak during our observations (in scales of tens of minutes), which supports the variability scenario. We should simultaneously consider that the spectrum of XTE$\,$J1810$-$197 may not always be well described over wide frequency ranges by a single power law. This is consistent with the spectral turnover reported by \citet{dai19} and has been observed for other radio magnetars \citep{cam08, tor17}. A conclusive answer for the spectral shape between centimetre and millimetre wavelengths, which would enable testing the radio emission mechanism \citep[e.g. revealing spectral turn-ups,][]{mich78}, requires simultaneous multi-frequency observations.

The single pulsations at 150$\,$GHz allow an estimation of the brightness temperature of the emission region, enabling us to test whether the millimetre emission is still coherent. For a distance to XTE$\,$J1810$-$197 of $3.5\pm0.5\,\rm{kpc}$ \citep{minter08} and a typical detected 0.75$\,$Jy pulse, the brightness temperature at 150$\,$GHz is $T_{\rm B}>10^{17}\,\rm{K}$ \citep[see e.g.][]{lorkra05}. We set the value as a lower limit because the pulses are not resolved, and narrower pulses correspond (following a light travel-time argument) to smaller emitting regions and higher brightness temperatures. Even at the lower limit, this high brightness temperature rules out incoherent mechanisms for the production of the pulses \citep[e.g.][]{singal09}. We therefore conclude that the underlying mechanism that causes the emission at 150$\,$GHz must still maintain a level of coherency. Without detected single pulses, we cannot constrain the emission mechanism at 260 GHz. We hypothesise, however, that because the continuum-equivalent flux density and pulse profile are so similar in the two frequency bands, the 260 GHz emission is probably caused by the same emission process. Additional observations with higher time-resolution are encouraged. They might help to further constrain the coherence of the emission mechanism by detecting narrower pulses\footnote{Time scales for individual pulses at lower frequencies are $\lesssim1\,$ms \citep{maan19}.}, perhaps also at 1.15$\,$mm or even shorter wavelengths.

While XTE$\,$J1810$-$197 was detected in the 2$\,$mm band during its previous epoch of activity in 2006 \citep[at 144$\,$GHz, $\uplambda$=2.08$\,$mm,][]{cam07c}, this is the first detection of the source at 260$\,$GHz ($\uplambda$=1.15$\,$mm). The fact that we see again strong emission from XTE$\,$J1810$-$197 in the millimetre band, as in 2006, suggests that the radio emission process that has been reactivated now is the same as in the previous period of activity. In the 2$\,$mm band, however, the emission is $\text{about ten}$ times stronger than what was reported by \citet{cam07c}.

The results presented here demonstrate that XTE$\,$J1810$-$197 produces strongly pulsed emission in the millimetre band after its reactivation in December 2018. The intrinsic variability of the source means that XTE$\,$J1810$-$197 opens a new opportunity of investigating the properties of pulsar emission at short millimetre wavelengths, where only another radio magnetar and two canonical pulsars have been detected above $50\,$GHz \citep{tor15, tor17, morr97, tor17_phd, liu19_almavela}. The current intensity and activity of this magnetar encourage observations at even shorter wavelengths, including submillimetre to optical bands.


Finally, we remark that this is the first time that KID technology as used in NIKA2 is successfully applied to detect broadband pulsations from neutron stars in the millimetre wavelength regime. In the future, the sampling frequency of NIKA2 may be increased, allowing for observations of faster-spinning pulsars or even burst emission of the kind of fast radio bursts \citep[FRB, see e.g.][]{corchat19}. 


\section{Conclusions}

The KID technology has the capability of detecting pulsations from neutron stars in the millimetre band, as demonstrated by these results for the magnetar XTE$\,$J1810$-$197 with the NIKA2 camera installed at the IRAM 30m telescope. Similarly, KIDs may be applied at even shorter wavelengths to observe pulsars and transient emission. 

The detections of XTE$\,$J1810$-$197 at 150 and 260$\,$GHz ($\uplambda$=2.0 and 1.15$\,$mm, respectively) are the first detections of the magnetar above 144$\,$GHz, showing that the star is again emitting intense millimetre radiation after its reactivation in December 2018. The spectrum in the millimetre band is relatively flat, but steeper than what was seen at centimetre wavelengths. The brightness temperature of the individual pulsations supports the hypothesis that a coherent mechanism is the source of the millimetre emission.

\begin{acknowledgements}
We thank the anonymous referee, Clemens Thum and Carsten Kramer for their constructive comments on the manuscript, and the NIKA2 team for supporting the observations. This work is based on observations carried out with the IRAM 30-m telescope. IRAM is supported by INSU/CNRS (France), MPG (Germany) and IGN (Spain). Financial support by the European Research Council for the ERC SynergyGrant BlackHoleCam (ERC-2013-SyG, Grant Agreement no. 610058) is gratefully acknowledged.
\end{acknowledgements}

%
\bibliographystyle{aa} 
\bibliography{biblio} 
%


\end{document}